\RequirePackage{amsthm}
\documentclass[default,iicol]{sn-jnl}

\usepackage{graphicx}%
\usepackage{multirow}%
\usepackage{amsmath,amssymb,amsfonts}%
\usepackage{amsthm}%
\usepackage{mathrsfs}%
\usepackage[title]{appendix}%
\usepackage{xcolor}%
\usepackage{textcomp}%
\usepackage{manyfoot}%
\usepackage{booktabs}%
\usepackage{algorithm}%
\usepackage{algorithmicx}%
\usepackage{algpseudocode}%
\usepackage{listings}%
\usepackage{cellspace} 
\usepackage{capt-of}

\RequirePackage{fix-cm}

\newcommand{\eg}{\textit{e.g.}, }
\newcommand{\ie}{\textit{i.e.}, }

\newcommand{\apr}{\textit{a priori} }

\usepackage{amsmath,amssymb}

\usepackage{ulem}
\usepackage{xcolor}
 
 \newcommand{\Autoref}[1]{%
  \begingroup%
  \def\chapterautorefname{Chapter}%
  \def\sectionautorefname{Section}%
  \def\subsectionautorefname{Subsection}%
  \autoref{#1}%
  \endgroup%
}
 
\usepackage{ulem}
\newcommand{\pvec}[1]{\vec{#1}\mkern2mu\vphantom{#1}}

\begin{document}
\title{Model bias and parameter optimisation with the example of the combination of nuclear models INCL and ABLA}

 \author*[1,2]{J.~Hirtz}\email{jason.hirtz@cea.fr}  
 \author[1]{J.-C.~David}  
 \author[2]{I.~Leya}  
 \author[3]{J.L.~Rodríguez-Sánchez}  
 \author[4]{G.~Schnabel}  

 \affil[1]{IRFU, CEA, Universit\'{e} Paris-Saclay, F-91191, Gif-sur-Yvette, France}
 \affil[2]{Space Research and Planetary Sciences, Physics Institute, University of Bern, Sidlerstrasse 5, 3012 Bern, Switzerland}
 \affil[3]{CITENI, Campus Industrial de Ferrol, Universidade da Coru\~{n}a, E-15403 Ferrol, Spain}
 \affil[4]{NAPC-Nuclear Data Section, International Atomic Energy Agency, Vienna, Austria}

\abstract{
 The accuracy and precision of high‑energy spallation models play a crucial role in the design and development of new applications and experiments, as well as in data analysis.
 We discuss the complementarity between parameter optimisation and model bias estimation approaches within a Bayesian framework.
 This is illustrated using the IntraNuclear Cascade model of Liège (INCL) together with the Ablation model (ABLA), for which these two approaches for model bias estimation have been applied independently in previous works.
 }
\maketitle

\section{Introduction}
\label{intro}

 The design, development, detection efficiency, and safety of experiments and facilities in nuclear physics require a large amount of reliable and consistent nuclear data.
 These data are also essential for interpreting complex experimental results during the data analysis phase and are notably employed in transport and activation models.  
 However, it is unrealistic to obtain all the mandatory data from experimental measurements; in particular, data above 20~MeV are rather scarce.
 Consequently, we rely heavily on model predictions to fill the gaps and to extrapolate into unmeasured regions.
 Thus, the accuracy and precision of models are essential for the aforementioned applications.
 
 Nowadays, precise models that deliver consistent and reliable results with well-defined uncertainties are essential across a variety of domains in order to advance both in fundamental and applied physics.
 Nonetheless, despite theoretical advances, nuclear models are far from reproducing trustworthy experimental data over wide ranges of observables and energies.
 
 Considering the need for reliable and consistent nuclear data on the one hand, and the difficulty of nuclear models to provide them on the other hand, we have developed two complementary approaches: one aims at improving the model predictions through optimised model parameters~\cite{opti} and another focuses on correcting the model bias~\cite{georg}.
 We applied these approaches to the IntraNuclear Cascade model of Liège (INCL)~\citep{inclBoudard,inclDavide,bibi} and to the Ablation model (ABLA)~\citep{JL2022,JL2023, JL2025}.
 Both approaches are based on Gaussian Process (GP) regression \cite{Rasmussen} and, by extension, belong to a Bayesian framework.

 In this paper, we explore the complementarity of these two approaches in order to provide accurate model predictions and reliable uncertainties. 
 
 The paper is organised as follows:
 First, we describe our objectives in \autoref{obj}.
 \Autoref{metho} develops the main ideas of parameter optimisation and model bias estimation.
 Then, the complementarity of the two approaches is discussed and illustrated with our results in \autoref{combi}.
 Next, we discuss the limits of the approach in \autoref{pb}.
 We also discuss the experimental data used for this study in \autoref{expSec}.
 Finally, we summarise our work in \autoref{conc}.
 
\section{Objectives}
\label{obj}
 
 The primary objective of this work is to enhance the accuracy and reliability of the INCL and ABLA models, which are widely used in both fundamental and applied nuclear physics and both models are integrated into major transport codes such as Geant4~\cite{geant4,geant41,geant42}.
 The specific goals of our approach are fourfold:

 \begin{itemize}
 \item[1] \textbf{Optimisation of model parameters:} An optimal set of parameters for the combination of INCL and ABLA models is determined, either in a general framework or for specific observables.
 In the former case, the optimal parameter set improves the accuracy of the model by better reproducing the underlying physics, thereby increasing precision.
 In the second case, the parameter optimisation can be used to investigate the physical processes related to the specific observable as illustrated in Ref.~\cite{opti} where we notably investigated the validity of strangeness production cross sections used in INCL.
 The parameter optimisation may flag missing or poorly represented processes in the code and indicate the direction for future model improvements if the optimised parameters are unrealistic.
 \item[2] \textbf{Estimation of parameter (joint) probability distributions:} The joint probability distribution quantifies the parameter uncertainties.
 The analysis of these distributions provides insights on how variations in model parameters propagate to model predictions and also shows how tightly the data constrain the parameters, leading to acceptable ranges for them.
 \item[3] \textbf{Model bias estimation:} Here we quantify the inherent bias of the INCL/ABLA model predictions, which allows us to correct it.
 Bias correction improves model accuracy, which directly impacts the design of instruments and the interpretation of experimental data.
 \item[4] \textbf{Post-correction uncertainty assessment:} After correcting for the bias, we provide ``systematic'' uncertainties on the model predictions.
 This provides a measure of confidence that enables users to make better/informed decisions based on the model output.
\end{itemize}

 The practical implications of achieving these objectives are substantial.
 Improved model accuracy and reduced uncertainty enhance our ability to design superior instrumentation and to interpret experimental results more comprehensively.
 These advances have broad applications, ranging from fundamental physics with the study of neutrinos, meteorites, antimatter, etc. to applied fields such as hadron therapy.
 By addressing the objectives outlined above, this research aims to deliver a more robust and reliable INCL/ABLA model combination, benefiting both theoretical investigations and practical applications in nuclear science and related disciplines.

\section{Methodology}
\label{metho}

 In previous publications~\citep{georg,opti}, we discussed two methods that improve the predictions of INCL/ABLA, as outlined in \autoref{obj}.  
 Here we recall the key concepts of these two methods.
 
 Firstly, it is worth emphasising that both approaches rely on the same underlying concept of GP regression~\cite{Rasmussen}.
 The formula used can be derived from Bayes' theorem under the assumption that all probability distributions are multivariate normal distributions ($\mathcal{N}$), as shown in Ref.~\cite{opti}.
 
 For two sets of observables $y_1$ and $y_2$, this hypothesis can be written as:
  \begin{equation}
  \rho(y_1,y_2) = \mathcal{N} \left(\begin{pmatrix}y_1 \\ y_2\end{pmatrix}\Bigg|\begin{pmatrix}\mu_1 \\ \mu_2\end{pmatrix}, 
  \begin{pmatrix} K_{11} & K_{12}\\ K_{21}& K_{22}\end{pmatrix}\right),
 \end{equation}
 with $\mu_1$ and $\mu_2$, the mean values of $y_1$ and $y_2$, respectively, and the $K_{ij}$ matrices, the sub-blocks of the covariance matrix $K$ between the elements of $y_1$ and/or $y_2$.
 
 The GP regression formula yields:
  \begin{equation}
  \rho(y_1|y_2) = \mathcal{N} \left(y_1 | \hat{y}_1  ,\hat{\Sigma}\right),
  \label{GP}
 \end{equation}
 with
  \begin{equation}
  \label{mu}
  \hat{y}_1=\mu_1 + K_{12} K_{22}^{-1} (y_2 - \mu_2),
 \end{equation}
  \begin{equation}
  \label{sig}
  \hat{\Sigma} = K_{11} - K_{12} K_{22}^{-1} K_{21}.
 \end{equation}
 Note that the hypothesis of a multivariate normal distribution implies a linear (but not deterministic) relation between $y_{1}$ and $y_{2}$, which one can find in \autoref{GP} and \autoref{mu}.
 
 In our framework, $y_1$ represents the observable(s) of interest we want to access, while $y_2$ denotes the experimental measurements.
 
 The methods we developed consist in applying \autoref{GP} by wisely replacing the abstract components $y$, $\mu$, and $K$ (\ie by selecting experimental data that are reproducible by the model and relevant to estimate the observable(s) of interest), by providing the priors $\mu_1$ and $\mu_2$, by evaluating the covariance matrix $K$, and by circumventing the limitations of the formula when the hypothesis of a multivariate normal distribution is not fulfilled.
 This is discussed in the following where the method is applied for both studies: parameter optimisation and model bias assessment.
 
\subsection{Parameter optimisation}
\label{paraOpti}
 
 The aim of the parameter optimisation is to improve the model predictions and the reliability of the model through a better representation of the microscopic features implemented within the model.
 The method used for the parameter optimisation is described with more details in Ref.~\cite{opti} and is summarised below.
 
 As mentioned above, we first need to define the components $y_1$ and $y_2$.
 Here, $y_1$, the observable of interest, is the optimal parameter set for the model $\pvec{p}_{\text{op}}$ and $y_2$, the experimental data, is a set $\vec{\sigma}_{\text{exp}}$ for which the corresponding model prediction is expected to depend on the model parameters to be optimised $\pvec{p}$.
 
 Next, priors for $y_1$ and $y_2$ must be provided.
 In the case of parameter optimisation, $\mu_1$ is our best \apr knowledge for the parameter set to be optimised.
 In practice, we use the original parameter values used in our models $\vec{p}_0$.
 $\mu_2$ corresponds to the model prediction for the observable $y_2$ using $\vec{p}_0$: $\mathcal{M}(\vec{p}_0)$.
 
 The construction of the covariance matrix $K$ is the most involved step because it requires the estimation of covariance matrices for the parameters $\Sigma_p$, for the experimental data $\Sigma_\text{exp}$, and for the correlation between the data and the parameters.
 The stochasticity of the model has to be taken into account as well.
 
 Here, $K_{11}$ is simply the covariance matrix of the parameters $\Sigma_p$ and is estimated based on an \apr knowledge.
 It is important to anticipate correlations when several parameters are involved in the description of the same feature within the model.
 
 $K_{12}$ links the uncertainties of the parameters to the observable uncertainties. It is given by the multiplication of the Jacobian (also called the sensitivity matrix) of the model evaluated in $\vec{p}_0$, hereafter noted as $J_{p_0}$, and the covariance matrix of the parameters:
 \begin{equation}
 K_{12} = K_{21}^T = \left(J_{p_0} \Sigma_p\right)^T = \Sigma_p J_{p_0}^T.
 \end{equation}
 
 $K_{22}$ can be divided into three contributions: the experimental covariance $\Sigma_{\text{exp}}$ (statistical + systematic uncertainties), the model prediction covariance arising from the stochasticity of the Monte‑Carlo simulation $\Sigma_{\text{sto}}$, and the correlation between the observables due to the dependence on the same parameters within the model.
 With this decomposition, the covariance matrix reads:
 \begin{equation}
 \label{sum_cov}
 K_{22} = \Sigma_{\text{exp}} + \Sigma_{\text{sto}} + J_{p_0} \Sigma_p J_{p_0}^T.
 \end{equation}
 In practice, if the statistics is good enough, this matrix $\Sigma_{\text{sto}}$ can be neglected.
 With this approach, the sources of model uncertainties other than the stochasticity (\ie the model bias) are supposed to be correctable by the choice of the optimal parameter set.
 In the following, we simplify $\Sigma_{\text{exp}} + \Sigma_{sto}$ by $\Sigma_e$ as these two matrices always sum up.
 This construction appears natural when deriving the GP formula as shown in Ref~\cite{opti}.
 
 Using Eqs.~\eqref{mu} and \eqref{sig} with the above definitions, the optimisation formulae become:
  \begin{equation}
   \label{eq_op_para}
   \pvec{p}_{\text{op}} = \vec{p}_0 + \Sigma_p J_{p_0}^T ( \Sigma_e + J_{p_0} \Sigma_p J_{p_0}^T )^{-1} (\vec{\sigma}_{\text{exp}} - \mathcal{M}(\vec{p}_0)),
 \end{equation}
 \begin{equation}
   \hat{\Sigma} = \Sigma_p - \Sigma_p J_{p_0}^T ( \Sigma_e + J_{p_0} \Sigma_p J_{p_0}^T )^{-1} J_{p_0} \Sigma_p.
   \label{sigma_para}
 \end{equation}
 
 Remember, these two equations assume that the model and the observables follow a multivariate normal distribution, an assumption that is never satisfied in practice as this would imply that the model is linear.
 However, the problem is alleviated by locally linearising the model with a Taylor expansion:
 \begin{equation}
 \label{Taylor}
 \vec{T}_i\left(\vec{p}\right) = \mathcal{M}\left(\vec{p}_i\right) + J_{p_i} (\vec{p}-\vec{p}_i),
 \end{equation}
 and by iterating Eqs.~\ref{eq_op_para} and \ref{sigma_para} with the update of $J_{p_0}$, which become $J_{p_i}$, and the update of $\mathcal{M}(\vec{p}_0)$, which become $T_i(\vec{p}_0)$, with the new parameter set $\vec{p}_{i}$ obtained in the previous loop.
 The convergence efficiency of this approach is directly linked to the linearity of the model.
 The closer the distribution $\rho(y_1,y_2)$ is to a multivariate normal distribution, the faster the convergence.
 \Autoref{eq_op_para} becomes:
  \begin{equation}
   \label{eq_para_iter}
   \pvec{p}_{i+1} = \vec{p}_0 + \Sigma_p J_{p_i}^T ( \Sigma_e + J_{p_i} \Sigma_p J_{p_i}^T )^{-1} (\vec{\sigma}_\text{exp} - T_i(\vec{p}_0)).
 \end{equation}
 
\begin{figure}[h]
   \centering
   \includegraphics[width=.95\columnwidth]{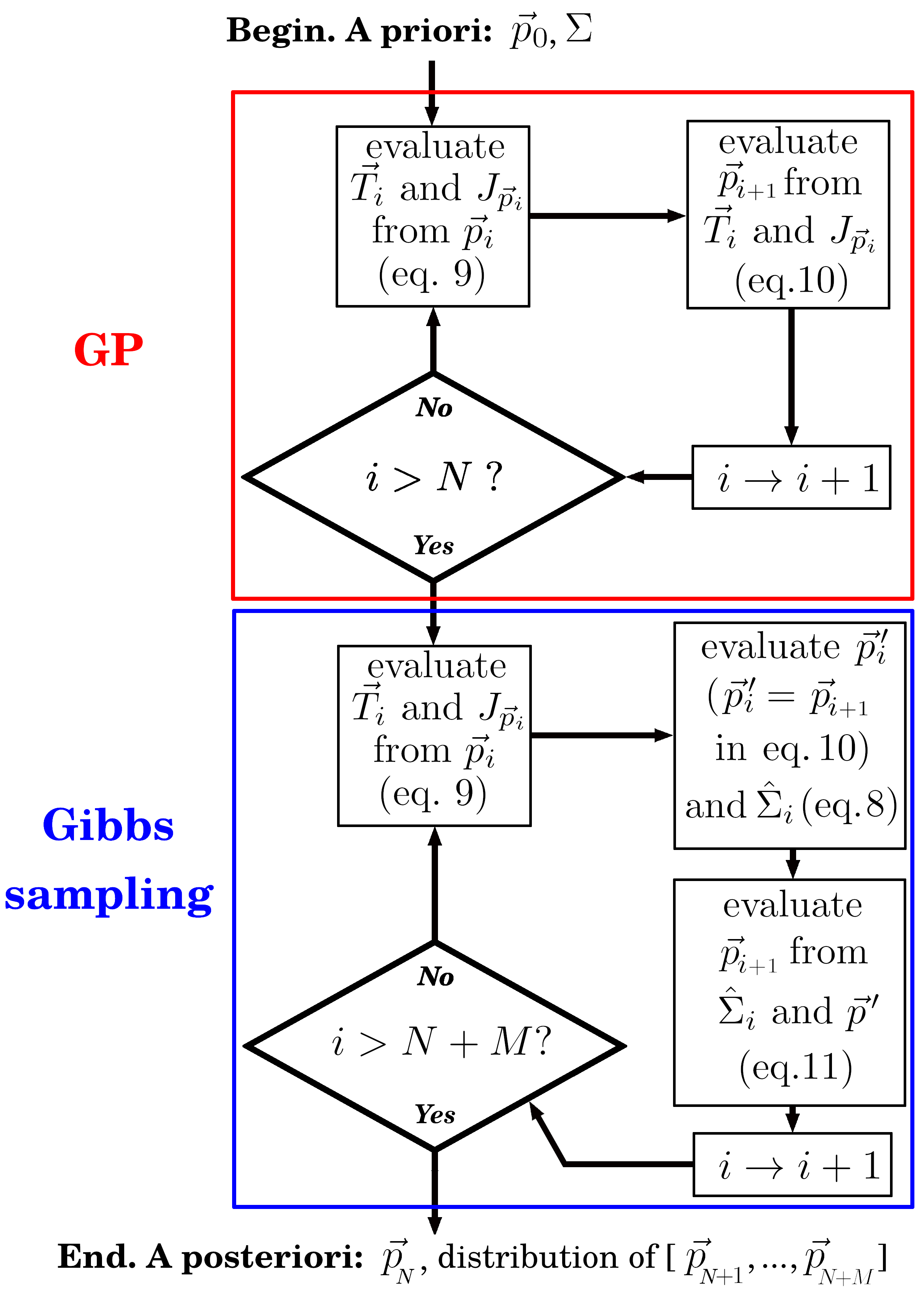}
   \caption{\label{loop} Flowchart representing the different steps of the algorithm allowing to pass from the prior $\vec{p}_0$ and its covariance $\Sigma$ to the posterior parameters $\vec{p}_N$ and the distribution $[\vec{p}_{N+1}$, ..., $\vec{p}_{N+M}]$.
   The latter can be turned into a posterior covariance matrix as described in the text.
   $N$ and $M$ are the numbers of iteration for the GP and Gibbs sampling, respectively.}
 \end{figure}

 The different steps of this first phase (hereafter called the GP phase) are illustrated in the upper part of \autoref{loop}.
 The number of iterations required depends on the observables studied, the model quality, the number of parameters to optimise, the Monte Carlo stochasticity, and the desired precision.
 In practice, we fix the number of iterations making the balance between these considerations and the implications on the running time (see \autoref{pb}).
 The number of iterations might range from a few tens to a hundred.
 The optimal parameter set is then obtained as the average of the parameter sets after the convergence (\ie the $\chi^2$ does not improve any more) in order to suppress the effect of the stochasticity of the Monte Carlo models.
 
 \autoref{sigma_para} can be used during the final iteration to obtain the updated uncertainties on the parameters.
 However, it requires the model likelihood to be at least roughly Gaussian around the optimal parameters to produce consistent uncertainties.
 By extension, the stochasticity must be negligible when evaluating the Jacobian as this can be seen as a source of non-linearity, which jeopardise the proper evaluation of $\hat{\Sigma}$.
 When this condition is not met, we need an additional phase to access the parameter uncertainties.
 One can have an indication that the model likelihood is not Gaussian if the dispersion of the parameter set after convergence is larger than the uncertainties obtained with $\hat{\Sigma}$.
 Note that this phase can be skipped if one is not interested in the parameter uncertainties, but only in the optimal parameters for the model.
 
 For this additional phase, we adopted a scheme that can be regarded as an approximation to Gibbs sampling \cite{gibbs}.
 The objective of the approach is to sample the distribution of the parameters, which then will provide uncertainties for the parameters.
 During this phase, we alternatively estimate the Jacobian $J_p$ from the parameter set $\vec{p}$ and sample $\vec{p}$ in the distribution $\pi(\vec{p}|J)$.
 However, we are not able to sample directly from $\pi(\vec{p}|J)$ because of the non-linear nature of the model which is not expressible in an analytic form.
 Therefore, for each iteration, we rely on a local linearisation of the model ($T_i$), which would result in a multivariate normal likelihood and therefore, in a parameter prior in which one can sample:
 \begin{equation}
 \vec{p}_{i+1} = \mathcal{N}( \pvec{p}'_i, \hat{\Sigma}_i),
 \end{equation}
 with $\pvec{p}'_i$ replacing $\vec{p}_{i+1}$ in \autoref{eq_para_iter}.
 Then, non-linearity of the model and the stochasticity in the Jacobian estimate are accounted for through the iterations with the re-evaluation of the Jacobian.
 This approach might be a source of bias if $\pi(\vec{p}|J)$ is too far from a multivariate normal distribution.
 However, for moderately non linear behaviour, the magnitude of posterior uncertainties and correlations can be evaluated with a reasonable number of iterations.
 The different steps of the Gibbs sampling-like approach are illustrated in the bottom part of \autoref{loop}.
 
 Ultimately, we have a series of parameter sets $\left[p_1, ..., p_{N+M}\right]$, with $N$ the number of iterations for the GP phase and $M$ the number of iterations for the Gibbs sampling like phase.
 The optimal parameter set is obtained from an average of parameter sets $\left[p_c, ..., p_{N}\right]$, with $c$ the number of iterations needed to observe a convergence of $\chi^2$.
 The uncertainties of these parameters are given by the updated covariance matrix $\hat{\Sigma}$ at the end of loop $N$ if we do not want or need to run the Gibbs sampling like phase or by the distribution of parameter sets $\left[p_{N+1}, ..., p_{N+M}\right]$ otherwise.
 Examples of applications are available in Ref.~\cite{opti} and in \autoref{combi}.

\subsection{Model bias estimation}
\label{bias}
 
 The second method we developed is dedicated to the evaluation of the model bias and is described in detail in Ref.~\cite{georg}.
 This method aims at evaluating model bias and at correcting the model output accordingly.
 It does not modify the model itself.
 The method is also based on \autoref{GP}.
 In order to apply the formula, we need once again to define the components.
 
 In this new case, $y_1$, the observable of interest, is the model bias at each location of interest and $\mu_1$ corresponds to the \apr bias of the model.
 Without evidence that the model is biased in a specific direction (\eg because of a missing feature with predictable effects), it is reasonable to use $\mu_1 = 0$ considering there is no reason to believe the model is more likely to overestimate than to underestimate the truth.
 Nevertheless, \apr bias can be added, without complicating the method.
 In the following, we assume $\mu_1 = 0$ to simplify the equations.
 
 For the definitions of $y_2$ and $\mu_2$, we can consider two possibilities, leading to the same result.
 One can consider $y_2$ as the ``experimental model bias'', in other words: the difference between the experimental data and the model, $\vec{\sigma}_\text{exp} - \mathcal{M}$, and $\mu_2$ would be the \apr model bias.
 In this case, one can use $\mu_2 = 0$ for the same reason mentioned for $\mu_1$.
 One can also consider $y_2$ as the set of experimental data, $\vec{\sigma}_\text{exp}$, and $\mu_2$ is our model prediction for the corresponding  observables corrected by our \apr model bias.
 This second definition corresponds to the one used for the parameter optimisation (without the \apr model bias, which is often zero in practice).
 It can simplify the comprehension when using the two methods together but it can also be a source of confusion: in parameter optimisation $\mu_2$ was a variable that changes with the parameter set along the iterations, whereas it is a constant here as we are not iterating.
 These two definitions do not modify the other components of the GP formula and will therefore lead to the same conclusion.
 Here, we will use the first definition to remain consistent with the notation used in Ref.~\cite{georg} and because it simplifies \autoref{mu} as:
 \begin{equation}
 \label{simp}
  \hat{y}_1=K_{12} K_{22}^{-1} y_2.
 \end{equation}
 
 Next, we need to define the covariance matrix $K$.
 As for \autoref{sum_cov}, there are three components in this matrix: the component of uncertainties introduced by the measurements, $\Sigma_\text{exp}$; the component of uncertainties introduced by the model, $\Sigma_\text{mod}$; and the component $\Sigma_\text{phys}$ describing the physical relation between the different observables:
 \begin{align} 
 K & = \begin{pmatrix} K_{11} & K_{12}\\ K_{21}& K_{22}\end{pmatrix} \\
   & = \begin{pmatrix} 0 & 0\\ 0 & \Sigma_\text{exp} \end{pmatrix} + \begin{pmatrix} 0 & 0\\ 0 & \Sigma_\text{mod} \end{pmatrix} + \Sigma_\text{phys} \nonumber
 \end{align}
 
 The same as for the parameter optimisation, the $\Sigma_\text{exp}$ contains the information about the uncertainties introduced by the measurements (statistic and systematic uncertainties) and it is accessed through the experimental publications and the databases (\eg EXFOR~\cite{exfor}, JENDL~\cite{jendl}, JEFF~\cite{jeff}, ENDF~\cite{endf}, etc.).
 The statistical model uncertainties in $\Sigma_\text{mod}$ can be calculated.
 These two parts are well constrained, although experimental systematic uncertainties can be debated because there are often unknowns.
 On the other hand, the uncertainties due to the model deficiency (\ie uncertainties other than those related to parameters) must be evaluated and incorporated in $\Sigma_\text{mod}$ in contrast to the parameter optimisation case, where we hypothesise that the model will be corrected by selecting an optimal parameter set.
 Another difference is that the physical relation between the observables cannot be accessed through the Jacobian of the model, as it was in \autoref{sum_cov}.
 The evaluation of $\Sigma_\text{phys}$ and the model deficiency in $\Sigma_\text{mod}$ will be achieved using Marginal Likelihood Optimisation (MLO).
 In our case, we used the L‑BFGS‑B algorithm \cite{L-BFGS-B} as implemented in the \textit{optim} function of R \cite{R}.
 In order to execute this algorithm, the structure of the covariance matrix has to be defined first, as it is not realistic to evaluate individually the correlation between each observable.
 The mathematical ``shape'' of the covariance matrix is commonly called a kernel.
 One can use various kernels to construct the missing part of the covariance matrix.
 The most common kernel is the square exponential covariance function:
 \begin{equation}
 \label{se}
 \kappa(x_i, x_j)= \exp \left(-\frac{(x_i-x_j)^2)}{2\lambda^2} \right).
 \end{equation}
 Here, the $\kappa(x_i, x_j)$'s are the elements of the covariance matrix $K$.
 There exists a large variety of kernels.
 Other common kernels are the periodic kernel, the Matérn covariance function, the linear kernel, and the trivial diagonal kernel.  
 Base kernels can be interpreted in physical terms.
 For example, the diagonal kernel corresponds to independent observables like a white noise or statistical uncertainties.
 Kernels can be flexibly combined to create new kernels by addition or multiplication to account for complex physical processes:
 \begin{equation}
 \kappa_{1+2}(x_i, x_j) = \kappa_1(x_i, x_j) + \kappa_2(x_i, x_j),
 \end{equation}
 \begin{equation}
 \kappa_{1\times 2}(x_i, x_j)= \kappa_1(x_i, x_j) \times \kappa_2(x_i, x_j).
 \end{equation}
 The resulting kernel is guaranteed to be valid by the Schur product theorem \cite{Schur}.
 More details can be found in \autoref{kernel} and in Ref.~\citep{Rasmussen}.
 The choice of the kernel is done by hand as a function of the physical processes we want to account for.
 All these kernels are defined with hyperparameters (\eg $\lambda$ in \autoref{se}), which will be estimated using MLO.
 The MLO can be interpreted as the algorithm searching for the simplest covariance matrix able to explain the difference between the experimental data and the model considering the structure imposed to the covariance matrix.
 
 In the cases discussed below, we decided to work with three kernels that we add up in addition to the known parts of the covariance matrix (\ie the statistical uncertainties): the Matérn covariance function (to describe the relation between the different observables), the constant kernel (to describe potential systematic biases of the model), and the diagonal kernel (to account for uncorrelated model deficiencies).
 The decision to use the Matérn covariance function instead of the usual square exponential covariance function is based on the observation that the L-BFGS-B algorithm is more robust to fake correlations that can appear in experimental data using this kernel (see \autoref{kernel} for details).
 The diagonal kernel can also be used to detect underestimated experimental error bars.
 This remains an arbitrary choice of kernel; one could select another set of kernels and obtain similar results.
 
 Once the covariance matrix $K$ is obtained with MLO, one can apply \autoref{simp} to predict the model bias.
 The posterior $\hat{y}_1$ obtained can be used to correct the model bias and \autoref{sig} gives us systematic uncertainties on the model prediction after correcting for the bias.
 
 Illustrative cases can be found in Ref.~\cite{georg} and in \autoref{combi}.
 
 \begin{figure*}[h]
   \centering
   \includegraphics[width=1.95\columnwidth]{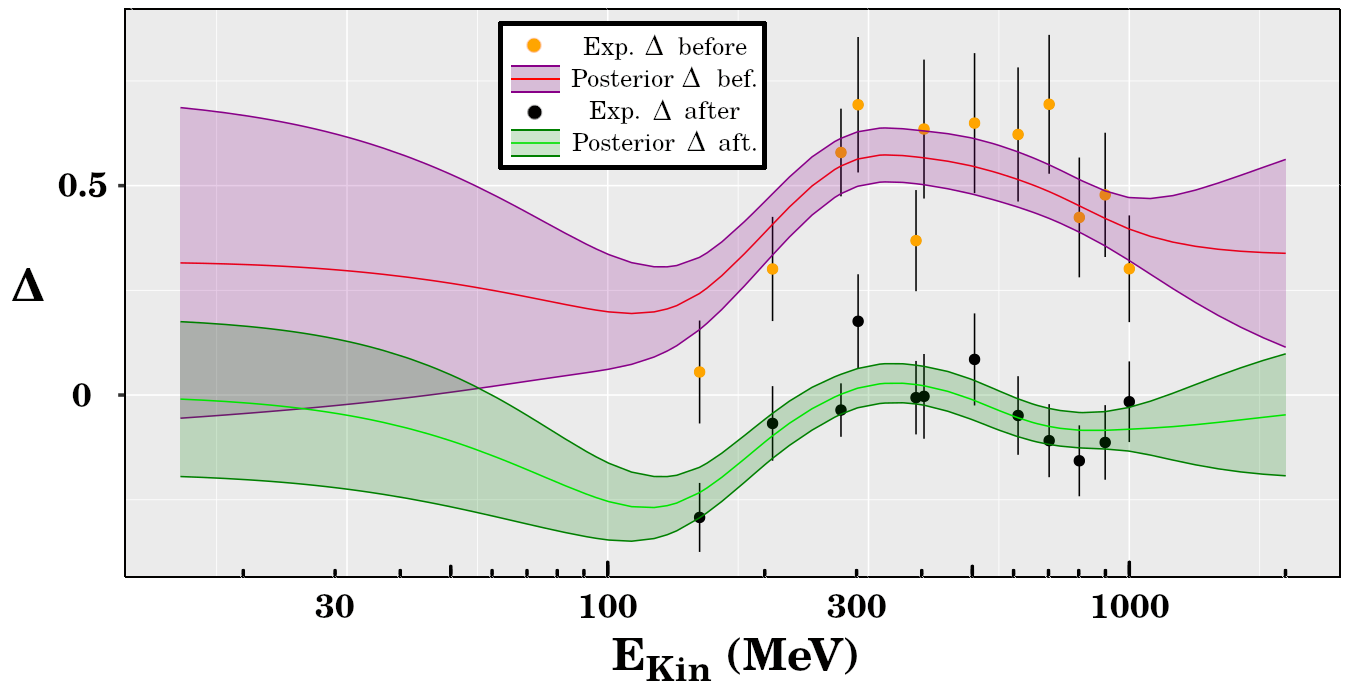}
   \caption{\label{beforeAfter} Estimations of the $\Delta$ quantity (see text) and its uncertainties before (red) and after (green) applying the parameter optimisation. The observable is the proton-induced fission cross section of $Bi$ as a function of the proton kinetic energy.}
 \end{figure*}
 
\section{Methods combination}
\label{combi}
 
 The two approaches described in \autoref{paraOpti} and \autoref{bias} are orthogonal in the sense that one can apply them independently.
 However, even though the two approaches play different roles, they exhibit a synergy.
 The parameter optimisation improves the model prediction by selecting the parameter set that is the most suitable for the model given the experimental data.
 This new parameter set can improve the model in two ways:
 it can arise from a better description of physical phenomena (the physical base is good but the original parameters are not) or from compensation of model deficiencies (the original parameters are correct but the physical base is not).
 In the first case, one can expect an improved coherence over a larger phase-space (energy, angles, ejectiles, etc.).
 This go together with the bias estimation, as more consistent predictions enhance the accuracy of bias assessment and reduce model uncertainty.
 However, the reduction of uncertainties is not guaranteed, because coherence may remain unchanged after the parameter improvement.
 In this case, there is no synergy, and the value of parameter optimisation is limited to a better understanding of the model (an example is available in Ref.~\cite{opti}). 
 In the second case, it is more difficult to conclude.
 However, it remains possible that compensation of model deficiencies lead to improved coherence over larger phase spaces than before, thereby yielding smaller model uncertainties.
 
 The interest in the methods combination is illustrated in \autoref{beforeAfter}.
 In this illustrative case of proton induced fission, we first evaluated the model bias of the INCL/ABLA model combination, intentionally deteriorating it to better visualise the effect of the method combination.
 The deterioration was achieved by modifying the fission‑dissipation coefficient in the ABLA model, fixing its initial value to $7(\pm5)$ instead of the default $4.5$ (arbitrary unit).
 
 The parameter optimisation algorithm was then applied to this fission‑dissipation coefficient and to the level density curvature.
 After the optimisation, the level density curvature parameter was slightly modified ($-0.016\pm 0.028$ for an initial $0\pm 0.1$) and the fission‑dissipation coefficient went back to $4.40 \pm 0.59$.
 With these new values, the $\chi^2/DoF$ decreased, passing from $4.8$ to $1.7$.
 As a remark, the two parameters also exhibit a strong correlation with a value of $0.96$.
 This means an increase in the level density curvature can compensate, within a certain range, an increase in the fission‑dissipation coefficient.
 
 Here we define the quantity $\Delta$, which is directly related to the relative model bias, as:
 \begin{equation}
 \Delta = \frac{\sigma-\sigma_{\text{mod}}}{\sigma_{\text{mod}}},
 \end{equation}
 where $\sigma$ is the value of an observable and $\sigma_\text{mod}$ its corresponding model prediction.
 An estimated $\Delta$ of $0.5$ would mean we should increase the model prediction by $50$\%, whereas a value of $-0.2$ would mean we should decrease it by $20$\%.
  
 In a first step, we display the evaluated $\Delta$ and its uncertainty for the proton induced cross section of Bi as a function of the proton kinetic energy for the model before optimising the parameters (in red and purple respectively in \autoref{beforeAfter}).
 The model bias estimation used experimental data for the proton induced cross section on Ho, Ta, Au, Pb, Bi, Th, U, Np, Pu 
\citep{Ayyad2014, Becchetti1983, Kruger1955, Bochagov1978, Yurevich2005, Bychenkov1973, Vaishnene2011, Flerov1972, JLRS2014, Schmidt2013, Konshin1966, Kotov1974, Smirnov1991, Stevenson1958, Ohtsuki1991, Yurevich2002, Isaev2008}.
 One can see that the model bias and its uncertainties are properly evaluated with respect to the experimental data.
 Compared to the same estimation after parameter optimisation (in green in \autoref{beforeAfter}), the improvement of the bias estimation in not obvious: it is properly evaluated in both cases.
 Indeed, we reproduce the experimental data with a $\chi^2/DoF < 1$ in each case after correcting for the estimated bias.
 Depending on the nuclei studied, the lowest $\chi^2$ can be obtained for the prediction before or after the parameter optimisation.
 However, the uncertainty on the bias estimation is reduced after the parameter optimisation, especially away from the experimental data.
 This means that we can have greater confidence in the model predictions over a wider range, which is a significant improvement over the approach evaluating the bias alone without parameter optimisation.
 
\section{Limits of the approach}
\label{pb}
 
 The method described here is very powerful and one can assume it could play a crucial role in the near future for a large range of applications requiring an accurate estimation of uncertainties.
 However, it is important to highlight the three main limitations to the use of these methods:
 
 \begin{itemize}
 \item[1] \textbf{The quality of the experimental data} used is the most limiting factor to the procedure because including automatically a large number of experimental data sets into the Bayesian procedure always bears the risk that some data sets are assigned too small uncertainties.
 Underestimated experimental uncertainties lead to a very high importance attributed to the associated experimental data and therefore to a biased result.
 Therefore, a careful study of the experimental data that are included in the Bayesian procedure must be carried out in order to use realistic or, at least, consistent uncertainties.
 Additionally, experimental data coming from a single experiment or using a similar set-up have common sources of systematic errors.
 These correlations must be included in the experimental covariance matrix $\Sigma_\text{exp}$, especially if the amount of data varies significantly from one dataset to another.
 This would prevent the attribution of a too high importance to an experiment with a large amount of highly correlated data versus another experiment with just a few data points.
 Further considerations about the construction of $\Sigma_\text{exp}$ are given in \autoref{expSec}.
 \item[2] \textbf{The computational power required} for this approach may be a real limiting factor for several independent reasons.
 First, the method requires inverting the $K_{22}$ matrix, for which the CPU complexity increases with $N^3$, with $N$ the number of experimental data points.
 Moreover, this procedure must be iterated for the parameter optimisation phase, which make it even more computing-time costly.
 Second, the model running time might be an issue, especially during the parameter optimisation phase as one must evaluate the code and its Jacobian iteratively.
 Last, the hyperparameter evaluation may also be CPU intensive for the model bias estimation phase with, once again, the inversion of the $K_{22}$ matrix within an iterative loop of the $optim$ function of R \cite{optimR}.
 With a modern desktop computer, one can apply the method up to a few thousand data points.
 
 Some technique were developed to reduce the running time of the method.
 In Ref.~\cite{georg}, we discussed a method developed in Ref.~\cite{pseudo}, where we define a matrix based on pseudo-input points instead of the real data points.
 These pseudo inputs are determined by a gradient based optimisation in such a way that the amount of pseudo-inputs $M$ is highly reduced with respect to the amount of real data points $N$.
 The training cost to obtain the pseudo inputs is in $\mathcal{O}(M^2N)$ \cite{pseudo}, which makes this approach relevant for the largest experimental data sets.
 Alternatively, approaches using small subsets of the experimental data exist.
 These approaches are called Sparse Gaussian Processes.
 The selection of the subset can be done based on information criterion \cite{Seeger}, or by changing the subset along the iterations of the algorithm \cite{Rasmussen}.
 
 Another technique we use to speed up the method is the reuse of the Jacobian of the model over several iterations along the parameter optimisation phase as it does not change drastically between two following iterations.
 The evaluation of the Jacobian requires to run many times the model with small variations on the parameter set used.
 Namely, we need to run the model $P+1$ times with $P$ the number of parameter to optimise.
 Therefore, the gain of time can be significant when the model running time is long.
 We implement a quick test after each iteration, which controls whether the Jacobian is still valid by comparing the prediction of the Taylor approximation with real model predictions.
 If the two predictions match within a predefined tolerance, one can continue with the same Jacobian.
 Otherwise, we re-evaluate the Jacobian in the following loop.
 
 \item[3] \textbf{The covariance matrix definition} is the last limitation to the approach.
 The covariance matrix $K$ used during the model bias estimation phase has a direct influence on the final estimation and its uncertainties.
 This is why it must be as reliable as possible.
 The use of the MLO will find the most reliable covariance matrix given the experimental data and considering the structure imposed by the user.
 If the structure proposed fails to capture the main correlations between the observables, the conclusion will be incorrect.
 On the other hand, a complex structure for the covariance matrix increases the probability of falling into a local minimum leading to an unrealistic covariance matrix.
 \end{itemize}

\section{Experimental data}
\label{expSec}
 
 As mentioned in previous sections, the information coming from the experimental data is central in the method.
 The experimental covariance matrix $\Sigma_\text{exp}$ contains the information about the statistical and systematic uncertainties.
 This matrix must be constructed properly.
 Otherwise, the result of our approach ($\hat{y}_1$ and $\hat{\Sigma}$) will be biased.
 However, a perfectly well defined $\Sigma_\text{exp}$ is unrealistic.
 A method as been developed in Ref.~\cite{georg2} to estimate $\Sigma_\text{exp}$ by comparing different sets of experimental data for the same observable.
 However, this approach can be applied only if several data sets exist, what is often not the case.
 
 In order to make this method applicable in a general case, we need to provide a ``non-unrealistic'' matrix $\Sigma_\text{exp}$.
 By ``non-unrealistic'', we mean that the $\Sigma_\text{exp}$ matrix must be realistic enough to account for the major source of uncertainties.
 This must include what is usually called the known unknowns (the reported statistics and systematics) and the unknown unknowns (the unthoughts sources on uncertainties) in the uncertainty quantification field.
 By nature, the known unknowns are easily accessible as they are reported with the experimental data.
 However, the unknown unknowns must be estimated based on our expertise.
 
 In the present study, we limited ourself to experimental data below 1.2~GeV because higher energies imply experimental set-ups in direct kinematic, which makes it very difficult to separate fission reactions from others.
 The data used here allow us to define $\Sigma_\text{exp}$ as a diagonal matrix without becoming too unrealistic.
 
\section{Summary and outlook}
\label{conc}
 
 In this work, we revisit the main objectives and features of the two independent approaches that we developed in previous studies about the parameter optimisation \cite{opti} and the model bias estimation \cite{georg} in the specific case of INCL coupled to the deexcitation model ABLA.
 In this article, we illustrate the possible synergy between these two approaches by applying them to the specific case of proton induced fission.
 We showed that the bias estimation approach alone is valid, even when the model is highly deficient, but the predictions are improved if the parameter optimisation approach is applied first.
 
 The parameter optimisation approach brought the model into better agreement with the experimental data.
 This increased our confidence in the extrapolation capabilities of INCL/ABLA and, by extension, reduced the uncertainties in the model predictions after correcting for the  bias.
 
 On the other hand, we highlighted the limitations that arise when the two approaches are combined.
 In particular, we emphasised the limitations related to the covariance matrix of the experimental data.
 In this work, we used a simple covariance matrix for the experimental data.
 In future, it may be possible to improve this matrix and, by extension, our results through in-depth study of the experimental data.
 
 We also proposed a covariance matrix kernel composed of the sum of a Matérn covariance function, a constant kernel, and a diagonal kernel.
 This kernel captures most of the possible correlations between observables while remaining resilient to fake correlations in the experimental data.
 
 These results are very encouraging, as they demonstrate our ability to provide accurate and precise model predictions with INCL/ABLA over a wide range of observables.
  
\section{Acknowledgment}
 This work is part of the NuRBS  project funded  by the french Agence Nationale de la Recherche under reference ANR-23-CE31-0008 and the Swiss National Science Foundation under the reference 200021E\_219157.
 J.L.~R.-S. acknowledges support from the \textit{Ram\'{o}n y Cajal} program under Grant No.~RYC2021-031989-I, funded by MCIN/AEI/10.13039/501100011033 and by the ``European Union NextGenerationEU/PRTR''.

\begin{appendices}

\section{Covariance matrix kernel}
\label{kernel}
 
\begin{table*}
\caption{\label{tab} List of standard (normalised) kernel considered in this study.}
\begin{tabular}{|Sc|Sc|Sc|Sc|}
\hline
Kernel name & Formula & Hyperparam. & Meaning \\
\hline
\hline
Diagonal & $\delta(x_i-x_j)$ & $\emptyset$ & White noise\\
\hline
Constant & 1 & $\emptyset$ & Systematic error\\
\hline
Square exponential & $\exp\left(-\frac{(x_i-x_j)^2}{2\lambda^2}\right)$ & $\lambda$ & Smooth variation\\
\hline
Matérn ($\nu = 3/2$) & $\left(1+\frac{\sqrt{3}|x_i-x_j|}{\rho}\right) \exp\left(-\frac{\sqrt{3}|x_i-x_j|}{\rho}\right)$ & $\rho$ & Smooth variation\\
\hline
Rational quadratic & $\left(1+\frac{(x_i-x_j)^2}{2\sigma^2\alpha}\right)^{-\alpha}$ & $\sigma$, $\alpha$ & Smooth variation\\
\hline
Periodic & $\exp\left(-\frac{2  \sin^2\left(\frac{\pi |x_i-x_j|}{k} \right)}{l^2}\right)$ & $k$, $l$ & Periodic correlation\\
\hline
\end{tabular}
\end{table*}
 
 A crucial question when estimating the model bias is the value of the covariance matrix $K$.
 In this appendix, we explain how the MLO is used to construct the covariance matrix when estimating the model bias.
 In particular, we discuss the different contributions and their impact on the covariance matrix.
 
 The use of the MLO to construct the covariance matrix requires defining the structure of this matrix.
 It is done through the use of various kernels (see below).
 An important feature of covariance matrices is that the independent contributions sum up \cite{Rasmussen}.
 Therefore, we need to find a kernel able to describe each independent contribution, and the structure of the covariance matrix will simply be the sum of all these kernels.
 
 In our study, we consider three types of contributions to the covariance matrix.
 First, the statistical uncertainties, which concern experimental data and the model prediction.
 This type of contribution is described by a diagonal matrix.
 
 Second, we have systematic uncertainties.
 This type of contribution corresponds to a shift affecting all the data in the same way.
 This is represented by a constant matrix ($\kappa(x_i,x_j)=c$).
 
 In the third type of contribution, we consider every other type of contribution.
 In practice, it corresponds to the physical correlations between the observables.
 In other words, it tells the ability to predict an observable knowing another one.
 For example, if the physics (\eg the total cross section $\sigma(p+p)$) is smooth within a certain energy domain, the fact that you know the value of the observable for a given energy (\eg 1~GeV) largely constrains the possible value for a nearby observation (\eg at 1.01~GeV) but this constraint rapidly weakens for observables further away (\eg at 10~GeV).
 Here, a large variety of kernels can be used to describe all the possible correlations between observables, depending on the physics studied.
 
 \begin{figure*}[h]
   \centering
   \includegraphics[width=1.9\columnwidth]{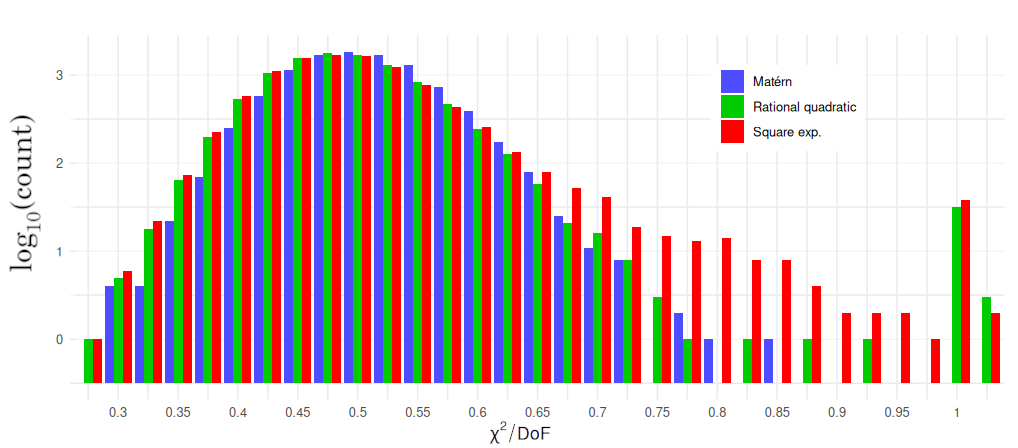}
   \caption{\label{histo} Distributions of the $\chi^2/DoF$ when using the different kernels in the construction of the covariance matrix.}
 \end{figure*}
 
 In \autoref{tab}, a list of basic (normalised) kernels we considered for the construction of the covariance matrix in the MLO algorithm is summarised, together with their hyperparameters and their concrete meaning in the covariance matrix.
 Other basic kernels exist, and more complex kernels can be constructed by multiplying two kernels \cite{Rasmussen}.
 Notably, non-stationary kernels exist.
 These kernels depend explicitly on $x_i$ and $x_j$, and not only on the difference $x_i-x_j$.
 This type of kernel can be used to describe a change of regime between two or more domains.
 Here, we focused on standard kernels for the sake of simplicity.
 
 While the contributions from statistical and systematic uncertainties are simple, the physical links between observables are much more complex.
 Many kernels can be used to describe the same phenomenon.
 In \autoref{tab}, one can mention the square exponential, the Matérn covariance, and the rational quadratic functions; all of which describe the smoothness of the physics with only minor differences.
 There is also the question of unexpected correlations.
 When choosing the structure of the covariance matrix for the MLO, we would like a versatile structure that can adapt to various situations, including unanticipated sources of correlation, while requiring the fewest possible hyperparameters to be adjusted, thereby avoiding overfitting and minimising the CPU cost of the method.
 
 In order to rank the best kernel structure to be used in our approach, we generated 40 data points from a given covariance matrix.
 Namely, the covariance matrix was the sum of three kernels: a diagonal kernel to represent statistical errors, a square exponential function for the smoothness of the curve, and a periodic kernel to represent an unanticipated term in the covariance function that cannot be reproduced by the other base kernels.
 The covariance matrix structure used for the MLO on these data consisted of a diagonal kernel plus one of the three kernels for the smoothness of the curve displayed in \autoref{tab}: the square exponential, the rational quadratic, and the Matérn ($\nu = 3/2$) function.
 This allows to test the ability of the structure to fit data without knowing the true structure used to generate the data.
 Note that the square exponential function has an advantage, as it appears explicitly in the ``true'' covariance matrix.
 
 \autoref{histo} presents the generalised $\chi^2/DoF$ distributions in our test case when the structure of the covariance matrix used for the MLO is based on the square exponential, the rational quadratic, or the Matérn ($\nu = 3/2$) function.
 The distributions were evaluated by repeating the MLO 10,000 times for each setup.
 We tested different sets of hyperparameters for the initial covariance matrix, including a case without the periodic contribution, but the following observations were always valid.
 As seen in \autoref{histo}, the different kernels perform very similar in average but the Matérn covariance function presents less extreme cases.
 In other words, the Matérn covariance function is less likely to produce unreasonable results.
 We observed that the rational quadratic function underperforms when falling into a local minimum during the optimisation, a situation that occurs more often than for other tested kernels because it involves more hyperparameters to optimise, resulting in $\chi^2/DoF \geq 1$ in our configuration.
 We explain the difference of performance between the Matérn ($\nu = 3/2$) covariance function and the square exponential function by the heavier tail for the Matérn ($\nu = 3/2$) covariance function.
 This makes the Matérn ($\nu = 3/2$) covariance function more resilient to fake correlations in the data on larger scales.
 This fake correlations correspond to similar shapes appearing in the data but for unrelated reasons.
 
 In conclusion, we decided to always use the same base for the covariance matrix: a diagonal kernel for the statistical uncertainties, a constant kernel for the systematics, and a Matérn covariance function to cover the other sources of covariance.
 We do not exclude the possibility to add other kernels in specific cases when it is motivated.
 
\end{appendices}
 
\bibliographystyle{unsrt} 
\bibliography{paper} 

\end{document}